\pdfoutput=1

\documentclass[11pt,twoside,a4paper,cmspaper,final,collab]{cms-tdr}
\def\svnVersion{473119P}\def\cmsCernNoTag{CERN-EP-2018-076}\def\cmsCernDate{\today}\def\cmsMessage{Published in Physical Review Letters as \href{http://dx.doi.org/10.1103/PhysRevLett.121.082301}{\doi{10.1103/PhysRevLett.121.082301.}}}

\begin{document}\cmsNoteHeader{HIN-17-003}

\hyphenation{had-ron-i-za-tion}
\hyphenation{cal-or-i-me-ter}
\hyphenation{de-vices}
\RCS$HeadURL: svn+ssh://svn.cern.ch/reps/tdr2/papers/HIN-17-003/trunk/HIN-17-003.tex $
\RCS$Id: HIN-17-003.tex 469350 2018-07-22 12:21:37Z alverson $
\newcommand{\pPb}{\ensuremath{\Pp\mathrm{Pb}}\xspace}
\newcommand {\PbPb}  {\ensuremath{\text{PbPb}}\xspace}
\newcommand{\pp}{\ensuremath{\Pp\Pp}\xspace}
\newcommand {\AuAu}  {\ensuremath{\text{AuAu}}\xspace}
\newcommand {\AonA}  {\ensuremath{\text{AA}}\xspace}
\newcommand {\dAu}    {\ensuremath{\text{dAu}}\xspace}
\newcommand{\noff}    {\ensuremath{N_\text{trk}^\text{offline}}\xspace}
\newcommand{\ket}       {\ensuremath{KE_{\mathrm{T}}}\xspace}
\newcommand {\deta}     {\ensuremath{\Delta\eta}}
\newcommand {\dphi}     {\ensuremath{\Delta\phi}}
\newcommand{\Dzerodecay}{\ensuremath{\PDz\to \PKm\Pgpp}\xspace}
\newcommand {\vnsub}    {\ensuremath{v_{2}^\text{sub}}}

\cmsNoteHeader{HIN-17-003}
\title{Elliptic flow of charm and strange hadrons in high-multiplicity \texorpdfstring{$\pPb$}{pPb} collisions at \texorpdfstring{$\sqrtsNN  = 8.16\TeV$}{sqrt(s[NN])  = 8.16 TeV}}

\date{\today}

\abstract{
The elliptic azimuthal anisotropy coefficient ($v_2$) is measured for
charm (\PDz) and strange (\PKzS, \PgL, \PgXm, and \PgOm) hadrons,
using a data sample of \pPb  collisions collected by the CMS experiment, at a nucleon-nucleon
center-of-mass energy of $\sqrtsNN  = 8.16\TeV$. A significant positive $v_2$
signal from long-range azimuthal correlations is observed for all
particle species in high-multiplicity \pPb collisions. The measurement
represents the first observation of possible long-range collectivity
for open heavy flavor hadrons in small systems.
The results suggest
that charm quarks have a smaller $v_2$ than the lighter quarks, probably
reflecting a weaker collective behavior. This effect is not seen in
the larger \PbPb collision system at $\sqrtsNN  = 5.02\TeV$, also presented.
}

\hypersetup{%
pdfauthor={CMS Collaboration},
pdftitle={Elliptic flow of charm and strange hadrons in high-multiplicity pPb collisions at sqrt(s[NN])  = 8.16 TeV},
pdfsubject={CMS},
pdfkeywords={CMS, physics, heavy ion, correlation, flow, pPb, heavy flavor, small system}}

\maketitle

There has been longstanding interest in the space-time evolution of the multiparticle production process in high energy collisions of hadrons~\cite{Busza:1975df}.
The observation of strong collective flow, as inferred from the correlations in azimuthal angle ($\phi$) of particles emitted
over a wide pseudorapidity ($\eta$) range in relativistic nucleus-nucleus (\AonA) collisions, has been one of the key signatures suggesting the formation of
a strongly interacting quark-gluon plasma (QGP) between the initial impact of the colliding nuclei and final production of particles.
As observed first at the BNL RHIC~\cite{Alver:2008gk,Adams:2005ph,Abelev:2009af,Alver:2008gk,Alver:2009id}
and later at the CERN LHC~\cite{Chatrchyan:2011eka,Chatrchyan:2012wg,Aamodt:2010pa,ATLAS:2012at,Chatrchyan:2012ta,CMS:2013bza},
the QGP exhibits nearly ideal hydrodynamical behavior~\cite{Ollitrault:1992bk,Heinz:2013th,Gale:2013da}.
In recent years, the observation of similar correlations
in events with high final-state particle multiplicity resulting from proton-proton
(\pp)~\cite{Khachatryan:2010gv,Aad:2015gqa,Khachatryan:2015lva}
and proton-lead (\pPb)~\cite{CMS:2012qk,alice:2012qe,Aad:2012gla,Aaij:2015qcq}
collisions at the LHC has raised the question whether a fluid-like QGP
is also created in these smaller collision systems~\cite{Weller:2017tsr}.

The azimuthal correlation structure of emitted particles is
typically characterized by its Fourier components~\cite{Voloshin:1994mz}. In hydrodynamic models,
the second and third Fourier components, known as elliptic ($v_2$) and triangular ($v_3$) flow, respectively,
most directly reflect the QGP response to the initial collision geometry and
its fluctuations~\cite{Alver:2010dn,Schenke:2010rr,Qiu:2011hf}.
The properties of the long-range correlation associated with light-flavor and
strange hadrons in small systems are found to be similar to those observed in \AonA\ collisions.
This includes, e.g., the particle species dependence~\cite{ABELEV:2013wsa,Khachatryan:2014jra,Khachatryan:2016txc}
and multiparticle (or collective) nature~\cite{Khachatryan:2015waa,Khachatryan:2016txc,Aaboud:2017acw,Aaboud:2017blb}
of the long-range correlation.
More recently, such long-range correlations have also been observed in lighter
systems at RHIC, including \dAu~\cite{Adare:2014keg,Adamczyk:2015xjc,Aidala:2017ajz}
and $^{3}$HeAu~\cite{Adare:2015ctn}.
While these measurements are consistent with a hydrodynamic expansion, alternative scenarios
based on gluon saturation in the initial state also claim to
capture the main features of the correlation data (recent reviews are provided in Refs.~\cite{Dusling:2015gta,Nagle:2018nvi}).

The large masses of heavy quarks (charm and bottom) lead to their being produced in the early stages of the collision,
and they thus probe the properties of the QGP
through their interactions with the medium~\cite{BraunMunzinger:2007tn}.
The elliptic flow results for \PD\ mesons in \AonA\ collisions measured
at RHIC~\cite{Adamczyk:2017xur} and the LHC~\cite{Abelev:2014ipa,Acharya:2017qps,Sirunyan:2017plt}
suggest that charm quarks develop strong collective behavior, similar to the bulk production of light flavor particles from the QGP.
In small systems,
long-range correlation involving inclusive muons and \JPsi mesons have revealed
hints of heavy flavor quark collectivity~\cite{Adam:2015bka,Acharya:2017tfn}.
Observation of \PD\ meson $v_2$ in the \pPb  system, and especially the comparison to the light-flavor and strange particle $v_2$,
can impose further constraints on different
interpretations related to the origin of the observed long-range collectivity.
In particular, such measurements can provide key insights to properties of heavy quark
interaction and thermalization within a hot QGP medium possibly formed
at a significantly reduced system size.

This Letter presents the first measurements of the elliptic anisotropies of prompt \PDz\ mesons
and strange hadrons
(\PKzS, \PgL, \PgXm, and \PgOm)
in \pPb collisions at a nucleon-nucleon center-of-mass energy of $\sqrtsNN  = 8.16\TeV$.
In all cases, particles and antiparticles are combined in the measurements.
The $v_2$ harmonic of all particle species in high-multiplicity
\pPb  events is determined in different intervals of transverse momentum (\pt),
via long-range two-particle correlations with charged particles.
The $v_2$ harmonics for the same strange hadrons are also measured in \PbPb
collisions at $\sqrtsNN  = 5.02\TeV$ for 30--50\% centrality (defined as the fraction of the total
inelastic cross section, with 0\% denoting the most central collisions), to compare
with the previously published \PDz\ $v_2$ results from these same collisions~\cite{Sirunyan:2017plt}.

The central feature of the CMS apparatus is a superconducting solenoid of 6\unit{m}
internal diameter, providing a magnetic field of 3.8\unit{T}. Within the solenoid volume,
there are four primary subdetectors including a silicon
pixel and strip tracker detector, a lead tungstate crystal electromagnetic calorimeter,
and a brass and scintillator hadron calorimeter, each composed of a
barrel and two endcap sections. Iron and quartz-fiber Cherenkov hadron forward
calorimeters cover the range $2.9 < \abs{\eta} < 5.2$.
Muons are measured in gas-ionization detectors
embedded in the steel flux-return yoke outside the solenoid. The silicon tracker
measures charged particles within the range $\abs{\eta}< 2.5$. For charged
particles with $1 < \pt < 10\GeV$ and $\abs{\eta} < 1.4$, the track resolutions are
typically 1.5\% in \pt and 25--90 (45--150)\mum in the transverse (longitudinal)
impact parameter~\cite{Chatrchyan:2014fea}. A detailed description of the CMS detector,
together with a definition of the coordinate system used and the relevant kinematic variables,
can be found in Ref.~\cite{Chatrchyan:2008zzk}.

The \pPb  data at $\sqrtsNN  = 8.16\TeV$ used in this analysis were
collected by the CMS experiment in 2016, and correspond to an integrated luminosity
of 186\nbinv. The beam energies are 6.5\TeV for the protons and
2.56\TeV per nucleon for the lead nuclei. Because of the asymmetric beam conditions,
particles selected in this Letter from midrapidity in the laboratory frame ($\abs{y_{\text{lab}}}<1$) correspond
to rapidity in the nucleon-nucleon center-of-mass frame of $-1.46 < y_{\text{cm}} < 0.54$,
with positive rapidity corresponding to the proton beam direction.
The event reconstruction, event selections, and triggers are identical to
those described in Refs.~\cite{Sirunyan:2017quh,Sirunyan:2017uyl}.
A subset of \PbPb  data at $\sqrtsNN = 5.02\TeV$ for 30--50\% centrality
is also used and reprocessed using the same reconstruction algorithm
as the \pPb  data.

The \pPb\ data are analyzed for multiplicity ranges of
$\noff < 35$ and $185 \leq \noff < 250$, where \noff\ is the number of
primary tracks~\cite{Chatrchyan:2014fea} with $\abs{\eta}<2.4$ and $\pt > 0.4\GeV$.
Events in the multiplicity region of $\noff > 250$ is not included to avoid effects
of multiple interactions in a single event (pileup).
These bin boundaries correspond to fractional inelastic cross sections from
100 to 57\%, and from 0.33 to 0.01\%, respectively.
The 30--50\% centrality \PbPb  data have an average \noff\ of 919.

The reconstruction and selection procedures for strange-hadron candidates in \pPb  collisions are
identical to those in Refs.~\cite{Khachatryan:2014jra,Khachatryan:2016yru,Khachatryan:2016txc}.
Pairs of oppositely charged particle tracks that
are detached from the primary vertex (i.e., having large significance of impact parameters) are selected to determine
if they point to a common secondary vertex from the decay of a \PKzS, a \PgL\ or a \PDz\ candidate.
The reconstruction efficiency for tracks with low momenta and large impact parameters
is increased by using all tracks that pass the loose selection of Ref.~\cite{Chatrchyan:2014fea}.
The track pair is assumed to originate from
$\Pgpp\Pgpm$ in \PKzS\ reconstruction, while the assumption of $\Pgpm\Pp\ (\Pgpp\Pap)$
is used in the \PgL\ (\PagL) reconstruction.
To reconstruct \PgXm\ and \PgOm\ particles, a \PgL\ candidate is combined with an additional particle
of the correct sign, assuming the pion or kaon PDG mass~\cite{PDG2016}, to define an additional vertex.
The \PDz\ mesons are reconstructed
through the hadronic decay channel \Dzerodecay.
In order to suppress the combinatorial background and improve the momentum and mass resolution,
high-purity~\cite{Chatrchyan:2014fea} tracks with $\pt > 0.7\GeV$ and $\abs{\eta} < 1.5$ are used.
For each pair of selected tracks, two \PDz\ candidates are considered by assuming
one of the tracks has the pion mass while the other track has the kaon mass, and vice versa.
Similar swapped-mass candidates are not required for strange hadrons since the two decay products are either the
same mass (for \PKzS) or so different in mass that momentum-sharing criteria can be used to identify which decay particle
is the heavier one (for \PgL, \PgXm\ and \PgOm).

Several topological selections are applied to further reduce the combinatorial background.
In particular, strange hadron and \PDz\ candidates are selected according to the
$\chi^2$ probability of their decay vertex, the three-dimensional distance (normalized by its uncertainty)
between the primary and decay vertices, and the pointing angle (defined as the angle between the line segment
connecting the primary and decay vertices and the momentum vector of the reconstructed particle candidates in the plane
transverse to the beam direction). The selection is optimized in each \pt bin, separately for different particle species, in order
to maximize the statistical significance of the signal.
For \PgXm\ and \PgOm\ reconstruction, these selections are applied to both the initial decay vertex and the subsequent
decay vertex of \PgL.

In the case of the \PDz\ measurement, the selections on the pointing angle also suppress the fraction
of nonprompt \PDz\ production (from decays of \PQb hadrons).
Simulated event samples of \PYTHIA 8.209~\cite{Sjostrand:2007gs,Khachatryan:2015pea} \PDz\ signal
events, embedded into {\textsc{epos lhc}}~\cite{Pierog:2013ria} minimum bias \pPb  events, are used
to estimate the nonprompt \PDz\ contamination in data.
By fitting the distributions of distance of closest approach of \PDz\ total momentum vector to the primary vertex,
using the probability distribution functions (pdf) for prompt and nonprompt
\PDz\ derived from simulation, the residual nonprompt fraction is found to be decreasing with \pt from 7 to 1\%.

The azimuthal anisotropies of \PDz\ mesons and strange hadrons
are extracted from their long-range ($\abs{\deta}>1$) two-particle azimuthal correlations with charged particles,
as described in Refs.~\cite{Khachatryan:2016txc,Khachatryan:2014jra}.
Taking the \PDz\ meson as an example, the two-dimensional (2D) correlation function
is constructed by pairing each \PDz\ candidate
with reference primary charged-particle tracks with $0.3<\pt<3.0\GeV$ (denoted ``ref'' particles), and calculating
\begin{linenomath}
\begin{equation}
\frac{1}{N_\text{\PDz}}\frac{\rd^{2}N^\text{pair}}{\rd\Delta\eta\, \rd\Delta\phi} = B(0,0)\ \frac{S(\Delta\eta,\Delta\phi)}{B(\Delta\eta,\Delta\phi)},
\end{equation}
\end{linenomath}
where $\Delta\eta$ and $\Delta\phi$ are the differences in $\eta$
and $\phi$ of each pair.
The same-event pair distribution, $S(\Delta\eta,\Delta\phi)$,
represents the yield of particle pairs normalized by the number of \PDz\
candidates from the same event. The mixed-event pair yield distribution,
$B(\Delta\eta,\Delta\phi)$,
is constructed by pairing \PDz\ candidates in each
event with the reference primary charged-particle tracks from 20 different randomly selected
events, from the same \noff\ range, and with a primary vertex falling in the same
2\unit{cm} wide range of reconstructed $z$ coordinate.
The analysis procedure is performed in each \PDz\ candidate \pt\ range by dividing it into intervals of invariant mass.
The correction for acceptance and efficiency (derived from {\textsc{pythia+epos}} simulations) of the \PDz\ meson yield
is found to have negligible effect on the measurements, and thus is not applied.
The \dphi\ correlation functions averaged over $\abs{\deta}>1$ (to
remove short-range correlations such as jet fragmentation) is then obtained from
the projection of 2D distributions and fitted by the first three terms of a Fourier series (including additional terms
has a negligible effect):
\begin{linenomath}
\begin{equation}
\label{eq:Vn}
\frac{1}{N_\text{\PDz}}\frac{\rd N^\text{pair}}{\rd\Delta\phi} = \frac{N_{\text{assoc}}}{2\pi} \left[ 1+\sum_{n=1}^{3} 2V_{n\Delta} \cos (n\Delta\phi)\right].
\end{equation}
\end{linenomath}
Here, $V_{n\Delta}$ are the Fourier coefficients and $N_{\text{assoc}}$
represents the total number of pairs per \PDz\ candidate.
By assuming $V_{n\Delta}$ to be the product of
single-particle anisotropies~\cite{Chatrchyan:2013nka}, $V_{n\Delta}(\PDz,{\text{ref}})=v_{n}(\PDz) v_{n}({\text{ref}})$,
the $v_n$ anisotropy harmonics for \PDz\ candidates can be extracted as a function of
invariant mass, $v_{n}(\PDz) = V_{n\Delta}(\PDz,{\text{ref}})/\sqrt{V_{n\Delta}({\text{ref}},{\text{ref}})}$.
\noindent Because of the limited amount of available data, only the elliptic
anisotropy harmonic is measured.
The residual contribution of back-to-back dijets to the measured $v_2$ results is corrected by subtracting correlations from
low-multiplicity \pPb  events, following an identical procedure established in Refs.~\cite{Chatrchyan:2013nka,Khachatryan:2016txc}.
The Fourier coefficients, $V_{n\Delta}$, extracted from events with $\noff < 35$ are subtracted from those extracted from
events with $185 \leq \noff < 250$ after accounting for the jet yield ratio of the selected events.
The subtraction is not performed for \PbPb  results as the back-to-back jet correlations are found
to be negligible in events with centrality between 30 and 50\%~\cite{Sirunyan:2017uyl}.

\begin{figure}[thb]
\centering
\includegraphics[width=\linewidth]{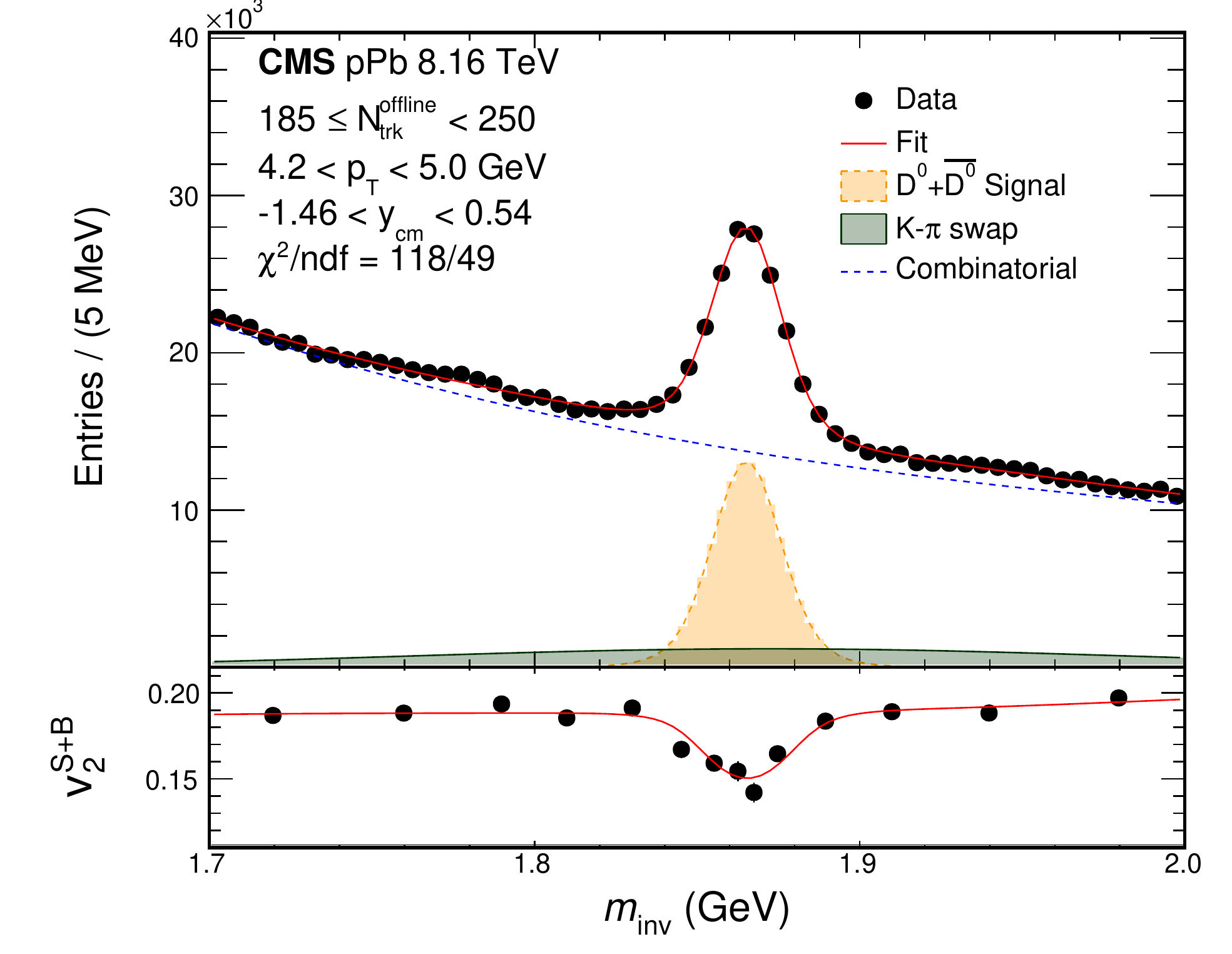}
\caption{Example of the simultaneous fit to the invariant mass spectrum
and $v_{2}^{S+B}(m_{\text{inv}})$ in the \pt interval 4.2--5.0\GeV for events with $185 \leq \noff < 250$.
}
\label{fig:mass_vn_fit}
\end{figure}

To extract the $v_{2}$ values of the \PDz\ meson signal ($v_{2}^{S}$),
a simultaneous fit to the invariant mass spectrum of \PDz\ candidates
and their $v_2$ as a function of the invariant mass, $v_{2}^{S+B}(m_{\text{inv}})$,
is performed in each \pt\ interval. The mass spectrum fit function is composed of
three components: the sum of two Gaussian functions with the same mean but different
widths for the \PDz\ signal, $S(m_{\text{inv}})$, an additional Gaussian function to describe the
invariant mass shape of \PDz\ candidates with an incorrect mass assignment from the
exchange of the pion and kaon designations, $SW(m_{\text{inv}})$, and a third-order polynomial
to model the combinatorial background, $B(m_{\text{inv}})$.
The width of $SW(m_{\text{inv}})$ and the ratio of the yields of $SW(m_{\text{inv}})$ and
$S(m_{\text{inv}})$ are fixed according to results obtained from {\textsc{pythia+epos}}
simulation studies.
The $v_{2}^{S+B}(m_{\text{inv}})$ distribution is fitted with
\begin{linenomath}
\begin{equation}
v_{2}^{S+B}(m_{\text{inv}}) = \alpha(m_{\text{inv}})v_{2}^{S} + [1-\alpha(m_{\text{inv}})]v_{2}^{B}(m_{\text{inv}}),
\end{equation}
\end{linenomath}
where
\begin{linenomath}
\begin{equation}
\alpha(m_{\text{inv}}) = \frac{S(m_{\text{inv}}) + SW(m_{\text{inv}})}{S(m_{\text{inv}}) + SW(m_{\text{inv}}) + B(m_{\text{inv}})}.
\end{equation}
\end{linenomath}
Here $v_{2}^{B}(m_{\text{inv}})$ for the background \PDz\ candidates is modeled as a linear function
of the invariant mass, and $\alpha(m_{\text{inv}})$ is the \PDz\ signal fraction. The \PK-\Pgp\
swapped component is included in the signal fraction because these candidates are from genuine \PDz\
mesons and should have the same $v_{2}$ value as that of the non-swapped \PDz\ signal.
Figure~\ref{fig:mass_vn_fit} shows an example of a simultaneous fit to the mass spectrum and $v_{2}^{S+B}(m_{\text{inv}})$ in
the \pt\ interval 4.2--5.0\GeV for the multiplicity range $185 \leq \noff < 250$ in \pPb  collisions.
The $v_2$ values for the strange hadrons are extracted in the same way although
no swapped-mass component is required.

\begin{figure}[htb]
\centering
\includegraphics[width=\linewidth]{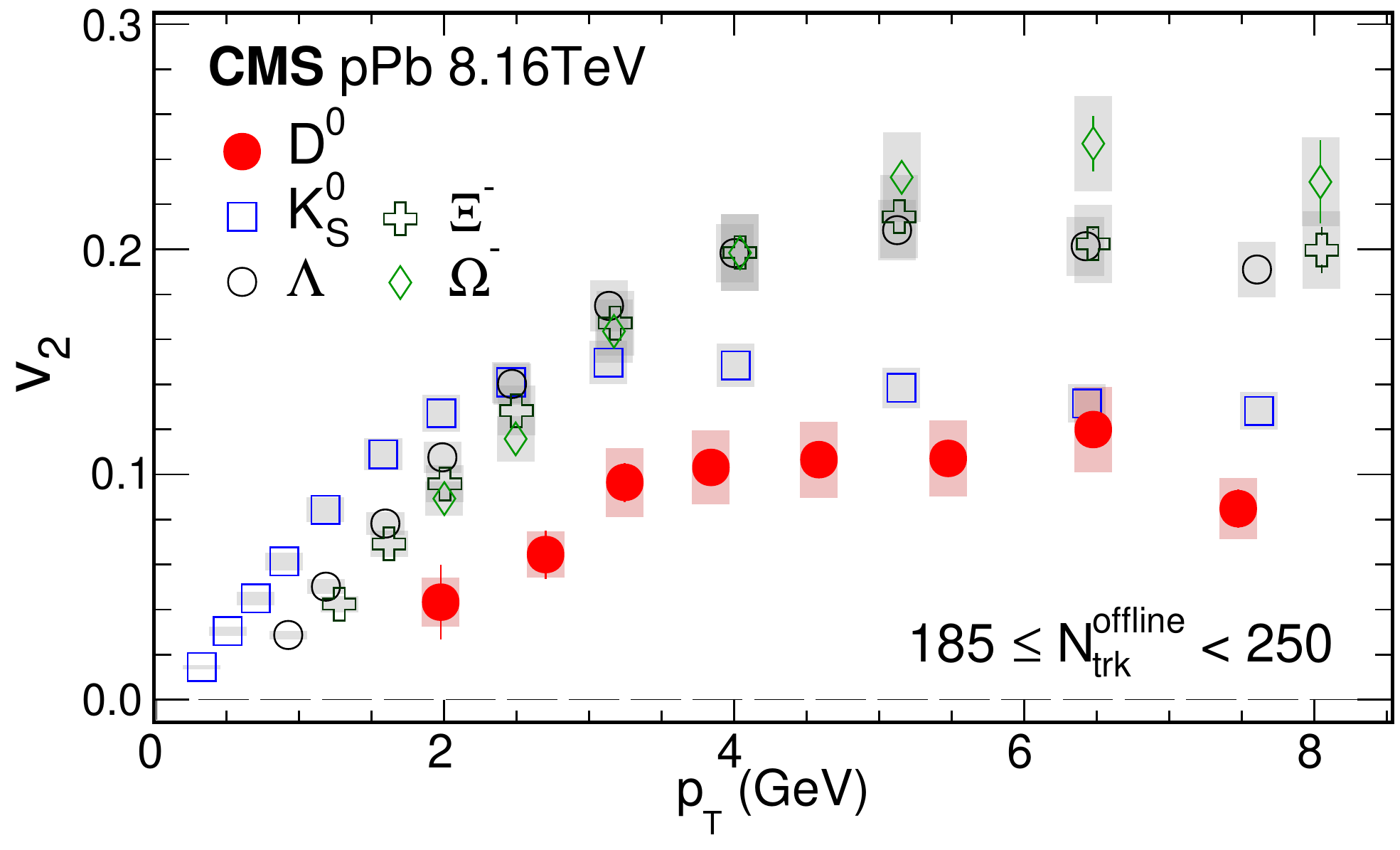}
\caption{Results of elliptic flow ($v_2$) for \PDz\ mesons, as well as \PKzS,
\PgL, \PgXm, and \PgOm\ particles, as functions of \pt\ for $-1.46 < y_{\mathrm{cm}} < 0.54$,
with $185 \leq \noff < 250$ in \pPb  collisions at $\sqrtsNN  = 8.16\TeV$.
The error bars correspond to statistical uncertainties, while the shaded areas denote
the systematic uncertainties.
 }
\label{fig:v2result}
\end{figure}

\begin{figure*}[thb]
\centering
\includegraphics[width=\linewidth]{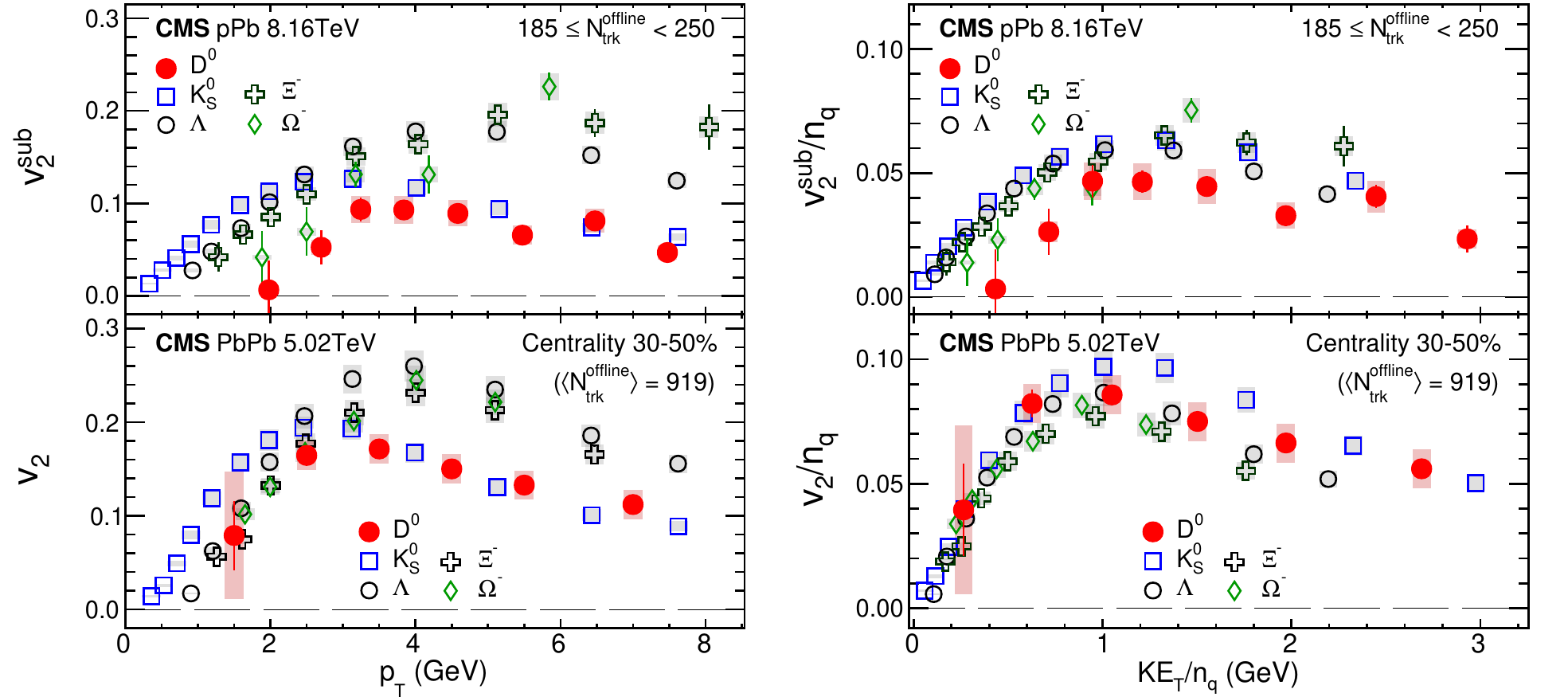}
\caption{
Left: the \vnsub\ results for prompt \PDz\ meson, as well as for strange hadrons,
as functions of \pt, for $-1.46 < y_{\mathrm{cm}} < 0.54$ in \pPb collisions at $\sqrtsNN  = 8.16\TeV$
with $185 \leq \noff < 250$ (upper) and for $-1 < y_{\mathrm{cm}} < 1$ in \PbPb  collisions at $\sqrtsNN  = 5.02\TeV$
for 30--50\% centrality (lower). Right: the $n_\cPq$-scaled \vnsub\ results.
The \PDz\ $v_2$ data in \PbPb  are taken from Ref.~\cite{Sirunyan:2017plt}.
The error bars correspond to statistical uncertainties, while the shaded areas denote
the systematic uncertainties.
 }
\label{fig:NCQsub}
\end{figure*}

As the residual contribution from nonprompt \PDz\ mesons is small,
no explicit correction is applied and a systematic uncertainty is quoted instead.
Based on the prediction for AA collisions that \PB\ mesons have a smaller $v_2$
than light-flavor particles, due to the larger mass of the b quark~\cite{Nahrgang:2014vza,He:2014cla,Xu:2015bbz},
the nonprompt \PDz\ $v_2$ values are assumed to lie between 0 and those of strange hadrons.
The maximum effect from nonprompt \PDz\ mesons is thus estimated using the extracted nonprompt \PDz\ fraction
and the change in $v_{2}^{S}$ is found to be smaller than 6\%.

Other sources of systematic uncertainty in the \PDz\ $v_2$ measurement in
this analysis include the background mass pdf, the \PDz\ meson yield correction
(acceptance and efficiency correction), selection of the \PDz\ candidates, and
the background $v_2$ pdf. No systematic effect has been observed while changing
the background mass pdf to a second-order polynomial or an exponential function.
To evaluate the uncertainties arising from the \PDz\ meson yield correction,
the $v_2$ values are extracted from the corrected signal \PDz\ distributions and compared to the
uncorrected $v_2$ values, yielding an uncertainty of 2\%.
The selection criteria for \PDz\ candidates are also varied to
tighter and looser values such that the \PDz\ signal fraction, $\alpha(m_{\text{inv}})$,
changes by 50\% and a systematic uncertainty of 14\% is evaluated from the variations of $v_{2}$.
The systematic uncertainties from the background $v_{2}$ pdf (20\% for $\pt < 2.4\GeV$ and 4\% for $\pt > 2.4 \GeV$)
are evaluated by changing $v_{2}^{B}(m_{\text{inv}})$ to a second-order polynomial function
of the invariant mass and a constant value. Systematic uncertainties from trigger bias and effects of pileup are negligible.

For \PKzS, \PgL, and \PgXm\ particles, the systematic uncertainties related to selection of reconstructed candidates
(2\% for \PKzS\ and \PgL\ particles and 6\% for \PgXm\ particles)
are evaluated in the same way as for \PDz\ mesons. To test the procedure of extracting
the signal $v_2$, a study using {\textsc{epos lhc}}~\cite{Pierog:2013ria} \pPb  events is performed
and the extracted values are compared to the generator-level values.
The agreement is found to be better than 6\%.
Systematic uncertainties for \PgOm\ particles are quoted to be the same as
those of \PgXm\ particles.

Figure~\ref{fig:v2result} shows the results of the $v_2$ measurement of the prompt \PDz\ meson
with $-1.46 < y_{\mathrm{cm}} < 0.54$ for high-multiplicity ($185 \leq \noff < 250$) \pPb  collisions.
The $v_2$ results for strange hadrons are also shown for comparison.
A clear mass ordering in the elliptic flow is observed in the low-\pt\ region of $\lesssim 2.5$\GeV,
where heavier particle species have a smaller $v_2$ signal at a given \pt\ value.
For $\pt > 2.5\GeV$, $v_2$ values for \PgL, \PgXm, and \PgOm\ baryons, which are similar to each other,
all become larger than those of \PDz\ and \PKzS\ mesons, a trend which is also observed in 5.02\TeV\
\pPb  collisions~\cite{Khachatryan:2014jra}.

The elliptic flow results corrected for residual jet correlations (\vnsub)
are shown in Fig.~\ref{fig:NCQsub} (upper left) for prompt \PDz\ mesons as well as
for strange hadrons as functions of
\pt\ for \pPb  collisions with $185 \leq \noff < 250$.
The reduction in the $v_2$ values resulting from the correction is most significant in the high-\pt\ region,
with a 30--40\% reduction found for $\pt > 5\GeV$.
The prompt \PDz\ \vnsub\ results show a clear trend of
rising and declining with \pt. The same ordering in particle mass as
seen in Fig.~\ref{fig:v2result} is observed in the \vnsub\ values in the lower-\pt\ region.
This behavior is consistent with the expectation of particle emission from a collective
expanding source,
which might indicate significant collective behavior of charm quarks
in high-multiplicity \pPb  systems at LHC energies.
Previously published $v_2$ data for \PDz\ meson in 30--50\% centrality
\PbPb  collisions~\cite{Sirunyan:2017plt}, together with new results for strange hadrons obtained in this Letter,
are shown in Fig.~\ref{fig:NCQsub} (lower left). A similar mass ordering to
that in \pPb  collisions is seen, although the multiplicity range is much larger.

Motivated by the quark coalescence model~\cite{Molnar:2003ff,Greco:2003xt,Fries:2003vb,Abelev:2007qg,Adare:2006ti},
collectivity at the partonic level is investigated by studying the scaling properties of \vnsub\
divided by the number of constituent quarks, $n_\cPq$, as a function of transverse kinetic energy per constituent quark,
$\ket/n_\cPq$ (where $\ket = \sqrt{\smash[b]{m^2 + \pt^2}} - m$),
for all hadronic species and systems measured (Fig.~\ref{fig:NCQsub}, right).
In high-multiplicity \pPb  collisions, the results for strange hadrons tend to follow
a universal trend in the region $0.5 < \ket/n_\cPq < 1.5\GeV$,
while the elliptic flow for \PDz\ mesons is found to have smaller values.
This could suggest that the collective behavior of charm quarks is weaker than that of the light-flavor
and strange quarks in high-multiplicity \pPb  collisions at the LHC. For $\ket/n_\cPq > 1.5\GeV$,
no clear universal scaling of $v_2/n_\cPq$ between mesons and baryons is observed.
The behavior is qualitatively different in the larger \PbPb  collision system with centrality between 30 and 50\%.
The results for all particle species tend to follow a common trend in the $\ket/n_\cPq < 1\GeV$ region,
indicating that \PDz\ mesons develop a strong collective behavior similar to the bulk of the QGP.

In summary, the first measurements of elliptic azimuthal anisotropies for prompt \PDz\ mesons,
as well as \PKzS, \PgL, \PgXm, \PgOm\ hadrons,
in high-multiplicity \pPb  collisions at $\sqrtsNN  = 8.16\TeV$ are presented. Significant
positive $v_2$ values are observed for \PDz\ mesons with $\pt > 2\GeV$. Comparing to strange-hadron results,
the \PDz\ $v_2$ values are found to be smaller at a given \pt, or at similar transverse kinetic energy per
constituent quark, after normalizing $v_2$ by the number of constituent quarks.
The latter effect is not observed in the larger \PbPb  collision system.
A possible interpretation is that, in high-multiplicity \pPb  collisions, in contrast to
larger nucleus-nucleus collision systems, the collective behavior of charm quarks is
weaker than that of the light-flavor quarks.

\begin{acknowledgments}
We congratulate our colleagues in the CERN accelerator departments for the excellent performance of the LHC and thank the technical and administrative staffs at CERN and at other CMS institutes for their contributions to the success of the CMS effort. In addition, we gratefully acknowledge the computing centers and personnel of the Worldwide LHC Computing Grid for delivering so effectively the computing infrastructure essential to our analyses. Finally, we acknowledge the enduring support for the construction and operation of the LHC and the CMS detector provided by the following funding agencies: BMWFW and FWF (Austria); FNRS and FWO (Belgium); CNPq, CAPES, FAPERJ, and FAPESP (Brazil); MES (Bulgaria); CERN; CAS, MoST, and NSFC (China); COLCIENCIAS (Colombia); MSES and CSF (Croatia); RPF (Cyprus); SENESCYT (Ecuador); MoER, ERC IUT, and ERDF (Estonia); Academy of Finland, MEC, and HIP (Finland); CEA and CNRS/IN2P3 (France); BMBF, DFG, and HGF (Germany); GSRT (Greece); NKFIA (Hungary); DAE and DST (India); IPM (Iran); SFI (Ireland); INFN (Italy); MSIP and NRF (Republic of Korea); LAS (Lithuania); MOE and UM (Malaysia); BUAP, CINVESTAV, CONACYT, LNS, SEP, and UASLP-FAI (Mexico); MBIE (New Zealand); PAEC (Pakistan); MSHE and NSC (Poland); FCT (Portugal); JINR (Dubna); MON, RosAtom, RAS and RFBR (Russia); MESTD (Serbia); SEIDI, CPAN, PCTI and FEDER (Spain); Swiss Funding Agencies (Switzerland); MST (Taipei); ThEPCenter, IPST, STAR, and NSTDA (Thailand); TUBITAK and TAEK (Turkey); NASU and SFFR (Ukraine); STFC (United Kingdom); DOE and NSF (USA).
\end{acknowledgments}

\bibliography{auto_generated}

\providecommand{\href}[2]{#2}\begingroup\raggedright\begin{thebibliography}{10}%
\makeatletter
\providecommand{\hrefCMSnoop }[0]{\@secondoftwo}%
\makeatother
\providecommand{\doi}{\texttt{doi:}\begingroup \urlstyle{tt}\Url}

\bibitem{Busza:1975df}
W.~Busza\hrefCMSnoop {}{ {et~al.}, ``{Charged particle multiplicity in
  \Pgpm-nucleus interactions at 100 and 175\GeVc.}'',} \textit{ Phys. Rev.
  Lett.} \textbf{ 34} (1975) 836,
\href{http://dx.doi.org/10.1103/PhysRevLett.34.836}{\doi{10.1103/PhysRevLett.34.836}}.

\bibitem{Alver:2008gk}
\hrefCMSnoop {}{{PHOBOS} Collaboration, ``{System size dependence of cluster
  properties from two- particle angular correlations in Cu+Cu and Au+Au
  collisions at \sqrtsNN\ = 200\GeV}'',} \textit{ Phys. Rev. C} \textbf{ 81}
  (2010) 024904,
  \href{http://dx.doi.org/10.1103/PhysRevC.81.024904}{\doi{10.1103/PhysRevC.81.024904}},
\href{http://www.arXiv.org/abs/0812.1172}{\texttt{arXiv:0812.1172}}.

\bibitem{Adams:2005ph}
\hrefCMSnoop {}{{STAR} Collaboration, ``{Distributions of charged hadrons
  associated with high transverse momentum particles in pp and Au+Au collisions
  at \sqrtsNN\ = 200\GeV}'',} \textit{ Phys. Rev. Lett.} \textbf{ 95} (2005)
  152301,
  \href{http://dx.doi.org/10.1103/PhysRevLett.95.152301}{\doi{10.1103/PhysRevLett.95.152301}},
\href{http://www.arXiv.org/abs/nucl-ex/0501016}{\texttt{arXiv:nucl-ex/0501016}}.

\bibitem{Abelev:2009af}
\hrefCMSnoop {}{{STAR} Collaboration, ``{Long range rapidity correlations and
  jet production in high energy nuclear collisions}'',} \textit{ Phys. Rev. C}
  \textbf{ 80} (2009) 064912,
  \href{http://dx.doi.org/10.1103/PhysRevC.80.064912}{\doi{10.1103/PhysRevC.80.064912}},
\href{http://www.arXiv.org/abs/0909.0191}{\texttt{arXiv:0909.0191}}.

\bibitem{Alver:2009id}
\hrefCMSnoop {}{{PHOBOS} Collaboration, ``{High transverse momentum triggered
  correlations over a large pseudorapidity acceptance in Au+Au collisions at
  \sqrtsNN\ = 200\GeV}'',} \textit{ Phys. Rev. Lett.} \textbf{ 104} (2010)
  062301,
  \href{http://dx.doi.org/10.1103/PhysRevLett.104.062301}{\doi{10.1103/PhysRevLett.104.062301}},
\href{http://www.arXiv.org/abs/0903.2811}{\texttt{arXiv:0903.2811}}.

\bibitem{Chatrchyan:2011eka}
\hrefCMSnoop {}{{CMS Collaboration}, ``{Long-range and short-range dihadron
  angular correlations in central PbPb collisions at a nucleon-nucleon center
  of mass energy of 2.76 TeV}'',} \textit{ JHEP} \textbf{ 07} (2011) 076,
  \href{http://dx.doi.org/10.1007/JHEP07(2011)076}{\doi{10.1007/JHEP07(2011)076}},
\href{http://www.arXiv.org/abs/1105.2438}{\texttt{arXiv:1105.2438}}.

\bibitem{Chatrchyan:2012wg}
\hrefCMSnoop {}{{CMS Collaboration}, ``{Centrality dependence of dihadron
  correlations and azimuthal anisotropy harmonics in PbPb collisions at
  \sqrtsNN\ = 2.76 TeV}'',} \textit{ Eur. Phys. J. C} \textbf{ 72} (2012) 2012,
  \href{http://dx.doi.org/10.1140/epjc/s10052-012-2012-3}{\doi{10.1140/epjc/s10052-012-2012-3}},
\href{http://www.arXiv.org/abs/1201.3158}{\texttt{arXiv:1201.3158}}.

\bibitem{Aamodt:2010pa}
\hrefCMSnoop {}{{ALICE Collaboration}, ``{Elliptic flow of charged particles in
  Pb-Pb collisions at 2.76 TeV}'',} \textit{ Phys. Rev. Lett.} \textbf{ 105}
  (2010) 252302,
  \href{http://dx.doi.org/10.1103/PhysRevLett.105.252302}{\doi{10.1103/PhysRevLett.105.252302}},
\href{http://www.arXiv.org/abs/1011.3914}{\texttt{arXiv:1011.3914}}.

\bibitem{ATLAS:2012at}
\hrefCMSnoop {}{{ATLAS Collaboration}, ``{Measurement of the azimuthal
  anisotropy for charged particle production in \sqrtsNN\ = 2.76\TeV lead-lead
  collisions with the ATLAS detector}'',} \textit{ Phys. Rev. C} \textbf{ 86}
  (2012) 014907,
  \href{http://dx.doi.org/10.1103/PhysRevC.86.014907}{\doi{10.1103/PhysRevC.86.014907}},
\href{http://www.arXiv.org/abs/1203.3087}{\texttt{arXiv:1203.3087}}.

\bibitem{Chatrchyan:2012ta}
\hrefCMSnoop {}{{CMS Collaboration}, ``{Measurement of the elliptic anisotropy
  of charged particles produced in PbPb collisions at nucleon-nucleon
  center-of-mass energy = 2.76 TeV}'',} \textit{ Phys. Rev. C} \textbf{ 87}
  (2013) 014902,
  \href{http://dx.doi.org/10.1103/PhysRevC.87.014902}{\doi{10.1103/PhysRevC.87.014902}},
\href{http://www.arXiv.org/abs/1204.1409}{\texttt{arXiv:1204.1409}}.

\bibitem{CMS:2013bza}
\hrefCMSnoop {}{{CMS Collaboration}, ``{Studies of azimuthal dihadron
  correlations in ultra-central PbPb collisions at \sqrtsNN\ = 2.76\TeV}'',}
  \textit{ JHEP} \textbf{ 02} (2014) 088,
  \href{http://dx.doi.org/10.1007/JHEP02(2014)088}{\doi{10.1007/JHEP02(2014)088}},
\href{http://www.arXiv.org/abs/1312.1845}{\texttt{arXiv:1312.1845}}.

\bibitem{Ollitrault:1992bk}
\hrefCMSnoop {}{J.-Y. Ollitrault, ``{Anisotropy as a signature of transverse
  collective flow}'',} \textit{ Phys. Rev. D} \textbf{ 46} (1992) 229,
\href{http://dx.doi.org/10.1103/PhysRevD.46.229}{\doi{10.1103/PhysRevD.46.229}}.

\bibitem{Heinz:2013th}
\hrefCMSnoop {}{U.~Heinz and R.~Snellings, ``{Collective flow and viscosity in
  relativistic heavy-ion collisions}'',} \textit{ Ann. Rev. Nucl. Part. Sci.}
  \textbf{ 63} (2013) 123,
  \href{http://dx.doi.org/10.1146/annurev-nucl-102212-170540}{\doi{10.1146/annurev-nucl-102212-170540}},
\href{http://www.arXiv.org/abs/1301.2826}{\texttt{arXiv:1301.2826}}.

\bibitem{Gale:2013da}
\hrefCMSnoop {}{C.~Gale, S.~Jeon, and B.~Schenke, ``{Hydrodynamic modeling of
  heavy-ion collisions}'',} \textit{ Int. J. Mod. Phys. A} \textbf{ 28} (2013)
  1340011,
  \href{http://dx.doi.org/10.1142/S0217751X13400113}{\doi{10.1142/S0217751X13400113}},
\href{http://www.arXiv.org/abs/1301.5893}{\texttt{arXiv:1301.5893}}.

\bibitem{Khachatryan:2010gv}
\hrefCMSnoop {}{{CMS Collaboration}, ``{Observation of long-range near-side
  angular correlations in proton-proton collisions at the LHC}'',} \textit{
  JHEP} \textbf{ 09} (2010) 091,
  \href{http://dx.doi.org/10.1007/JHEP09(2010)091}{\doi{10.1007/JHEP09(2010)091}},
\href{http://www.arXiv.org/abs/1009.4122}{\texttt{arXiv:1009.4122}}.

\bibitem{Aad:2015gqa}
\hrefCMSnoop {}{{ATLAS Collaboration}, ``{Observation of long-range elliptic
  azimuthal anisotropies in $\sqrt{s}=$13 and 2.76 TeV pp collisions with the
  ATLAS Detector}'',} \textit{ Phys. Rev. Lett.} \textbf{ 116} (2016) 172301,
  \href{http://dx.doi.org/10.1103/PhysRevLett.116.172301}{\doi{10.1103/PhysRevLett.116.172301}},
\href{http://www.arXiv.org/abs/1509.04776}{\texttt{arXiv:1509.04776}}.

\bibitem{Khachatryan:2015lva}
\hrefCMSnoop {}{{CMS Collaboration}, ``{Measurement of long-range near-side
  two-particle angular correlations in pp collisions at $\sqrt s =$13 TeV}'',}
  \textit{ Phys. Rev. Lett.} \textbf{ 116} (2016) 172302,
  \href{http://dx.doi.org/10.1103/PhysRevLett.116.172302}{\doi{10.1103/PhysRevLett.116.172302}},
\href{http://www.arXiv.org/abs/1510.03068}{\texttt{arXiv:1510.03068}}.

\bibitem{CMS:2012qk}
\hrefCMSnoop {}{{CMS Collaboration}, ``{Observation of long-range near-side
  angular correlations in proton-lead collisions at the LHC}'',} \textit{ Phys.
  Lett. B} \textbf{ 718} (2013) 795,
  \href{http://dx.doi.org/10.1016/j.physletb.2012.11.025}{\doi{10.1016/j.physletb.2012.11.025}},
\href{http://www.arXiv.org/abs/1210.5482}{\texttt{arXiv:1210.5482}}.

\bibitem{alice:2012qe}
\hrefCMSnoop {}{{ALICE Collaboration}, ``{Long-range angular correlations on
  the near and away side in \pPb\ collisions at \sqrtsNN\ = 5.02\TeV }'',}
  \textit{ Phys. Lett. B} \textbf{ 719} (2013) 29,
  \href{http://dx.doi.org/10.1016/j.physletb.2013.01.012}{\doi{10.1016/j.physletb.2013.01.012}},
\href{http://www.arXiv.org/abs/1212.2001}{\texttt{arXiv:1212.2001}}.

\bibitem{Aad:2012gla}
\hrefCMSnoop {}{{ATLAS Collaboration}, ``Observation of associated near-side
  and away-side long-range correlations in $\sqrt{s_{_{NN}}}=5.02$ tev
  proton-lead collisions with the {ATLAS} detector'',} \textit{ Phys. Rev.
  Lett.} \textbf{ 110} (2013) 182302,
  \href{http://dx.doi.org/10.1103/PhysRevLett.110.182302}{\doi{10.1103/PhysRevLett.110.182302}},
\href{http://www.arXiv.org/abs/1212.5198}{\texttt{arXiv:1212.5198}}.

\bibitem{Aaij:2015qcq}
\hrefCMSnoop {}{{LHCb Collaboration}, ``{Measurements of long-range near-side
  angular correlations in \sqrtsNN\ = 5\TeV proton-lead collisions in the
  forward region}'',} \textit{ Phys. Lett. B} \textbf{ 762} (2016) 473,
  \href{http://dx.doi.org/10.1016/j.physletb.2016.09.064}{\doi{10.1016/j.physletb.2016.09.064}},
\href{http://www.arXiv.org/abs/1512.00439}{\texttt{arXiv:1512.00439}}.

\bibitem{Weller:2017tsr}
\hrefCMSnoop {}{R.~D. Weller and P.~Romatschke, ``{One fluid to rule them all:
  viscous hydrodynamic description of event-by-event central p+p, p+Pb and
  Pb+Pb collisions at $\sqrt{s}=5.02$ TeV}'',} \textit{ Phys. Lett. B} \textbf{
  774} (2017) 351,
  \href{http://dx.doi.org/10.1016/j.physletb.2017.09.077}{\doi{10.1016/j.physletb.2017.09.077}},
\href{http://www.arXiv.org/abs/1701.07145}{\texttt{arXiv:1701.07145}}.

\bibitem{Voloshin:1994mz}
\hrefCMSnoop {}{S.~Voloshin and Y.~Zhang, ``{Flow study in relativistic nuclear
  collisions by Fourier expansion of azimuthal particle distributions}'',}
  \textit{ Z. Phys. C} \textbf{ 70} (1996) 665,
  \href{http://dx.doi.org/10.1007/s002880050141}{\doi{10.1007/s002880050141}},
\href{http://www.arXiv.org/abs/hep-ph/9407282}{\texttt{arXiv:hep-ph/9407282}}.

\bibitem{Alver:2010dn}
\hrefCMSnoop {}{B.~H. Alver, C.~Gombeaud, M.~Luzum, and J.-Y. Ollitrault,
  ``{Triangular flow in hydrodynamics and transport theory}'',} \textit{ Phys.
  Rev. C} \textbf{ 82} (2010) 034913,
  \href{http://dx.doi.org/10.1103/PhysRevC.82.034913}{\doi{10.1103/PhysRevC.82.034913}},
\href{http://www.arXiv.org/abs/1007.5469}{\texttt{arXiv:1007.5469}}.

\bibitem{Schenke:2010rr}
\hrefCMSnoop {}{B.~Schenke, S.~Jeon, and C.~Gale, ``Elliptic and triangular
  flow in event-by-event {D=3+1} viscous hydrodynamics'',} \textit{ Phys. Rev.
  Lett.} \textbf{ 106} (2011) 042301,
  \href{http://dx.doi.org/10.1103/PhysRevLett.106.042301}{\doi{10.1103/PhysRevLett.106.042301}},
\href{http://www.arXiv.org/abs/1009.3244}{\texttt{arXiv:1009.3244}}.

\bibitem{Qiu:2011hf}
\hrefCMSnoop {}{Z.~Qiu, C.~Shen, and U.~Heinz, ``{Hydrodynamic elliptic and
  triangular flow in Pb-Pb collisions at \sqrtsNN\ = 2.76\TeV}'',} \textit{
  Phys. Lett. B} \textbf{ 707} (2012) 151,
  \href{http://dx.doi.org/10.1016/j.physletb.2011.12.041}{\doi{10.1016/j.physletb.2011.12.041}},
\href{http://www.arXiv.org/abs/1110.3033}{\texttt{arXiv:1110.3033}}.

\bibitem{ABELEV:2013wsa}
\hrefCMSnoop {}{{ALICE Collaboration}, ``{Long-range angular correlations of
  pi, K and p in p--Pb collisions at \sqrtsNN\ = 5.02\TeV}'',} \textit{ Phys.
  Lett. B} \textbf{ 726} (2013) 164,
  \href{http://dx.doi.org/10.1016/j.physletb.2013.08.024}{\doi{10.1016/j.physletb.2013.08.024}},
\href{http://www.arXiv.org/abs/1307.3237}{\texttt{arXiv:1307.3237}}.

\bibitem{Khachatryan:2014jra}
\hrefCMSnoop {}{{CMS Collaboration}, ``{Long-range two-particle correlations of
  strange hadrons with charged particles in pPb and PbPb collisions at LHC
  energies}'',} \textit{ Phys. Lett. B} \textbf{ 742} (2015) 200,
  \href{http://dx.doi.org/10.1016/j.physletb.2015.01.034}{\doi{10.1016/j.physletb.2015.01.034}},
\href{http://www.arXiv.org/abs/1409.3392}{\texttt{arXiv:1409.3392}}.

\bibitem{Khachatryan:2016txc}
\hrefCMSnoop {}{{CMS Collaboration}, ``{Evidence for collectivity in pp
  collisions at the LHC}'',} \textit{ Phys. Lett. B} \textbf{ 765} (2017) 193,
  \href{http://dx.doi.org/10.1016/j.physletb.2016.12.009}{\doi{10.1016/j.physletb.2016.12.009}},
\href{http://www.arXiv.org/abs/1606.06198}{\texttt{arXiv:1606.06198}}.

\bibitem{Khachatryan:2015waa}
\hrefCMSnoop {}{{CMS Collaboration}, ``{Evidence for collective multi-particle
  correlations in pPb collisions}'',} \textit{ Phys. Rev. Lett.} \textbf{ 115}
  (2015) 012301,
  \href{http://dx.doi.org/10.1103/PhysRevLett.115.012301}{\doi{10.1103/PhysRevLett.115.012301}},
\href{http://www.arXiv.org/abs/1502.05382}{\texttt{arXiv:1502.05382}}.

\bibitem{Aaboud:2017acw}
\hrefCMSnoop {}{{ATLAS Collaboration}, ``{Measurement of multi-particle
  azimuthal correlations in $pp$, $p+$Pb and low-multiplicity Pb$+$Pb
  collisions with the ATLAS detector}'',} \textit{ Eur. Phys. J. C} \textbf{
  77} (2017) 428,
  \href{http://dx.doi.org/10.1140/epjc/s10052-017-4988-1}{\doi{10.1140/epjc/s10052-017-4988-1}},
\href{http://www.arXiv.org/abs/1705.04176}{\texttt{arXiv:1705.04176}}.

\bibitem{Aaboud:2017blb}
\hrefCMSnoop {}{{ATLAS Collaboration}, ``{Measurement of long-range
  multiparticle azimuthal correlations with the subevent cumulant method in pp
  and p+Pb collisions with the ATLAS detector at the CERN Large Hadron
  Collider}'',} \textit{ Phys. Rev. C} \textbf{ 97} (2018) 024904,
  \href{http://dx.doi.org/10.1103/PhysRevC.97.024904}{\doi{10.1103/PhysRevC.97.024904}},
\href{http://www.arXiv.org/abs/1708.03559}{\texttt{arXiv:1708.03559}}.

\bibitem{Adare:2014keg}
\hrefCMSnoop {}{{PHENIX} Collaboration, ``{Measurement of long-range angular
  correlation and quadrupole anisotropy of pions and (anti)protons in central
  $d$$+$Au collisions at \sqrtsNN\ = 200 GeV}'',} \textit{ Phys. Rev. Lett.}
  \textbf{ 114} (2015) 192301,
  \href{http://dx.doi.org/10.1103/PhysRevLett.114.192301}{\doi{10.1103/PhysRevLett.114.192301}},
\href{http://www.arXiv.org/abs/1404.7461}{\texttt{arXiv:1404.7461}}.

\bibitem{Adamczyk:2015xjc}
\hrefCMSnoop {}{{STAR} Collaboration, ``{Long-range pseudorapidity dihadron
  correlations in $d$+Au collisions at \sqrtsNN\ = 200 GeV}'',} \textit{ Phys.
  Lett. B} \textbf{ 747} (2015) 265,
  \href{http://dx.doi.org/10.1016/j.physletb.2015.05.075}{\doi{10.1016/j.physletb.2015.05.075}},
\href{http://www.arXiv.org/abs/1502.07652}{\texttt{arXiv:1502.07652}}.

\bibitem{Aidala:2017ajz}
\hrefCMSnoop {}{{PHENIX} Collaboration, ``{Measurements of multiparticle
  correlations in d+Au collisions at 200, 62.4, 39, and 19.6 GeV and p+Au
  collisions at 200 GeV and implications for collective behavior}'',} \textit{
  Phys. Rev. Lett.} \textbf{ 120} (2018) 062302,
  \href{http://dx.doi.org/10.1103/PhysRevLett.120.062302}{\doi{10.1103/PhysRevLett.120.062302}},
\href{http://www.arXiv.org/abs/1707.06108}{\texttt{arXiv:1707.06108}}.

\bibitem{Adare:2015ctn}
\hrefCMSnoop {}{{PHENIX} Collaboration, ``{Measurements of elliptic and
  triangular flow in high-multiplicity $^{3}$He$+$Au collisions at \sqrtsNN\ =
  200 GeV}'',} \textit{ Phys. Rev. Lett.} \textbf{ 115} (2015) 142301,
  \href{http://dx.doi.org/10.1103/PhysRevLett.115.142301}{\doi{10.1103/PhysRevLett.115.142301}},
\href{http://www.arXiv.org/abs/1507.06273}{\texttt{arXiv:1507.06273}}.

\bibitem{Dusling:2015gta}
\hrefCMSnoop {}{K.~Dusling, W.~Li, and B.~Schenke, ``{Novel collective
  phenomena in high-energy proton--proton and proton--nucleus collisions}'',}
  \textit{ Int. J. Mod. Phys. E} \textbf{ 25} (2016) 1630002,
  \href{http://dx.doi.org/10.1142/S0218301316300022}{\doi{10.1142/S0218301316300022}},
\href{http://www.arXiv.org/abs/1509.07939}{\texttt{arXiv:1509.07939}}.

\bibitem{Nagle:2018nvi}
\hrefCMSnoop {}{J.~L. Nagle and W.~A. Zajc, ``{Small system collectivity in
  relativistic hadron and nuclear collisions}'',} (2018).
  \href{http://www.arXiv.org/abs/1801.03477}{\texttt{arXiv:1801.03477}}.
Submitted to Ann. Rev. Nucl. Part. Sci.

\bibitem{BraunMunzinger:2007tn}
\hrefCMSnoop {}{P.~Braun-Munzinger, ``{Quarkonium production in
  ultra-relativistic nuclear collisions: Suppression versus enhancement}'',}
  \textit{ J. Phys. G} \textbf{ 34} (2007) S471,
  \href{http://dx.doi.org/10.1088/0954-3899/34/8/S36}{\doi{10.1088/0954-3899/34/8/S36}},
\href{http://www.arXiv.org/abs/nucl-th/0701093}{\texttt{arXiv:nucl-th/0701093}}.

\bibitem{Adamczyk:2017xur}
\hrefCMSnoop {}{{STAR} Collaboration, ``{Measurement of \PDz\ azimuthal
  anisotropy at midrapidity in Au+Au collisions at $\sqrtsNN = 200\GeV $}'',}
  \textit{ Phys. Rev. Lett.} \textbf{ 118} (2017) 212301,
  \href{http://dx.doi.org/10.1103/PhysRevLett.118.212301}{\doi{10.1103/PhysRevLett.118.212301}},
\href{http://www.arXiv.org/abs/1701.06060}{\texttt{arXiv:1701.06060}}.

\bibitem{Abelev:2014ipa}
\hrefCMSnoop {}{{ALICE Collaboration}, ``{Azimuthal anisotropy of D meson
  production in Pb-Pb collisions at $\sqrtsNN = 2.76\TeV $}'',} \textit{ Phys.
  Rev. C} \textbf{ 90} (2014) 034904,
  \href{http://dx.doi.org/10.1103/PhysRevC.90.034904}{\doi{10.1103/PhysRevC.90.034904}},
\href{http://www.arXiv.org/abs/1405.2001}{\texttt{arXiv:1405.2001}}.

\bibitem{Acharya:2017qps}
\hrefCMSnoop {}{{ALICE Collaboration}, ``{$D$-meson azimuthal anisotropy in
  midcentral Pb-Pb collisions at \sqrtsNN\ = 5.02 TeV}'',} \textit{ Phys. Rev.
  Lett.} \textbf{ 120} (2018) 102301,
  \href{http://dx.doi.org/10.1103/PhysRevLett.120.102301}{\doi{10.1103/PhysRevLett.120.102301}},
\href{http://www.arXiv.org/abs/1707.01005}{\texttt{arXiv:1707.01005}}.

\bibitem{Sirunyan:2017plt}
\hrefCMSnoop {}{{CMS Collaboration}, ``{Measurement of prompt D$^{0}$ meson
  azimuthal anisotropy in PbPb collisions at \sqrtsNN\ = 5.02\TeV}'',} (2017).
  \href{http://www.arXiv.org/abs/1708.03497}{\texttt{arXiv:1708.03497}}.
Submitted to Phys. Rev. Lett.

\bibitem{Adam:2015bka}
\hrefCMSnoop {}{{ALICE Collaboration}, ``{Forward-central two-particle
  correlations in p-Pb collisions at $\sqrt{s_{\rm NN}}$ = 5.02 TeV}'',}
  \textit{ Phys. Lett. B} \textbf{ 753} (2016) 126,
  \href{http://dx.doi.org/10.1016/j.physletb.2015.12.010}{\doi{10.1016/j.physletb.2015.12.010}},
\href{http://www.arXiv.org/abs/1506.08032}{\texttt{arXiv:1506.08032}}.

\bibitem{Acharya:2017tfn}
\hrefCMSnoop {}{{ALICE Collaboration}, ``{Search for collectivity with
  azimuthal J/$\psi$-hadron correlations in high multiplicity p-Pb collisions
  at $\sqrt{s_{\rm NN}}$ = 5.02 and 8.16 TeV}'',} \textit{ Phys. Lett. B}
  \textbf{ 780} (2018) 7,
  \href{http://dx.doi.org/10.1016/j.physletb.2018.02.039}{\doi{10.1016/j.physletb.2018.02.039}},
\href{http://www.arXiv.org/abs/1709.06807}{\texttt{arXiv:1709.06807}}.

\bibitem{Chatrchyan:2014fea}
\hrefCMSnoop {}{{CMS Collaboration}, ``{Description and performance of track
  and primary-vertex reconstruction with the CMS tracker}'',} \textit{ JINST}
  \textbf{ 9} (2014) P10009,
  \href{http://dx.doi.org/10.1088/1748-0221/9/10/P10009}{\doi{10.1088/1748-0221/9/10/P10009}},
\href{http://www.arXiv.org/abs/1405.6569}{\texttt{arXiv:1405.6569}}.

\bibitem{Chatrchyan:2008zzk}
\hrefCMSnoop {}{{CMS Collaboration}, ``The {CMS} experiment at the {CERN}
  {LHC}'',} \textit{ JINST} \textbf{ 3} (2008) S08004,
\href{http://dx.doi.org/10.1088/1748-0221/3/08/S08004}{\doi{10.1088/1748-0221/3/08/S08004}}.

\bibitem{Sirunyan:2017quh}
\hrefCMSnoop {}{{CMS Collaboration}, ``{Constraints on the chiral magnetic
  effect using charge-dependent azimuthal correlations in pPb and PbPb
  collisions at the LHC}'',} (2017).
  \href{http://www.arXiv.org/abs/1708.01602}{\texttt{arXiv:1708.01602}}.
Submitted to Phys. Rev. C.

\bibitem{Sirunyan:2017uyl}
\hrefCMSnoop {}{{CMS Collaboration}, ``{Observation of correlated azimuthal
  anisotropy fourier harmonics in pp and p+Pb collisions at the LHC}'',}
  \textit{ Phys. Rev. Lett.} \textbf{ 120} (2018) 092301,
  \href{http://dx.doi.org/10.1103/PhysRevLett.120.092301}{\doi{10.1103/PhysRevLett.120.092301}},
\href{http://www.arXiv.org/abs/1709.09189}{\texttt{arXiv:1709.09189}}.

\bibitem{Khachatryan:2016yru}
\hrefCMSnoop {}{{CMS Collaboration}, ``{Multiplicity and rapidity dependence of
  strange hadron production in pp, pPb, and PbPb collisions at the LHC}'',}
  \textit{ Phys. Lett. B} \textbf{ 768} (2017) 103,
  \href{http://dx.doi.org/10.1016/j.physletb.2017.01.075}{\doi{10.1016/j.physletb.2017.01.075}},
\href{http://www.arXiv.org/abs/1605.06699}{\texttt{arXiv:1605.06699}}.

\bibitem{PDG2016}
\hrefCMSnoop {}{{Particle Data Group}, C.~Patrignani {et~al.}, ``Review of
  particle physics'',} \textit{ Chin. Phys. C} \textbf{ 40} (2016) 100001,
  \href{http://dx.doi.org/10.1088/1674-1137/40/10/100001}{\doi{10.1088/1674-1137/40/10/100001}}.

\bibitem{Sjostrand:2007gs}
\hrefCMSnoop {}{{T. Sj\"ostrand, S. Mrenna and P. Skands}, ``{A brief
  introduction to PYTHIA 8.1}'',} \textit{ Comput. Phys. Commun.} \textbf{ 178}
  (2008) 852,
  \href{http://dx.doi.org/10.1016/j.cpc.2008.01.036}{\doi{10.1016/j.cpc.2008.01.036}},
\href{http://www.arXiv.org/abs/0710.3820}{\texttt{arXiv:0710.3820}}.

\bibitem{Khachatryan:2015pea}
\hrefCMSnoop {}{{CMS Collaboration}, ``{Event generator tunes obtained from
  underlying event and multiparton scattering measurements}'',} \textit{ Eur.
  Phys. J. C} \textbf{ 76} (2016) 155,
  \href{http://dx.doi.org/10.1140/epjc/s10052-016-3988-x}{\doi{10.1140/epjc/s10052-016-3988-x}},
\href{http://www.arXiv.org/abs/1512.00815}{\texttt{arXiv:1512.00815}}.

\bibitem{Pierog:2013ria}
T.~Pierog\hrefCMSnoop {}{ {et~al.}, ``{EPOS LHC: Test of collective
  hadronization with data measured at the CERN Large Hadron Collider}'',}
  \textit{ Phys. Rev. C} \textbf{ 92} (2015) 034906,
  \href{http://dx.doi.org/10.1103/PhysRevC.92.034906}{\doi{10.1103/PhysRevC.92.034906}},
\href{http://www.arXiv.org/abs/1306.0121}{\texttt{arXiv:1306.0121}}.

\bibitem{Chatrchyan:2013nka}
\hrefCMSnoop {}{{CMS} Collaboration, ``{Multiplicity and transverse momentum
  dependence of two- and four-particle correlations in \pPb\ and \PbPb\
  collisions}'',} \textit{ Phys. Lett. B} \textbf{ 724} (2013) 213,
  \href{http://dx.doi.org/10.1016/j.physletb.2013.06.028}{\doi{10.1016/j.physletb.2013.06.028}},
\href{http://www.arXiv.org/abs/1305.0609}{\texttt{arXiv:1305.0609}}.

\bibitem{Nahrgang:2014vza}
M.~Nahrgang\hrefCMSnoop {}{ {et~al.}, ``{Elliptic and triangular flow of heavy
  flavor in heavy-ion collisions}'',} \textit{ Phys. Rev. C} \textbf{ 91}
  (2015) 014904,
  \href{http://dx.doi.org/10.1103/PhysRevC.91.014904}{\doi{10.1103/PhysRevC.91.014904}},
\href{http://www.arXiv.org/abs/1410.5396}{\texttt{arXiv:1410.5396}}.

\bibitem{He:2014cla}
\hrefCMSnoop {}{M.~He, R.~J. Fries, and R.~Rapp, ``{Heavy flavor at the Large
  Hadron Collider in a strong coupling approach}'',} \textit{ Phys. Lett. B}
  \textbf{ 735} (2014) 445,
  \href{http://dx.doi.org/10.1016/j.physletb.2014.05.050}{\doi{10.1016/j.physletb.2014.05.050}},
\href{http://www.arXiv.org/abs/1401.3817}{\texttt{arXiv:1401.3817}}.

\bibitem{Xu:2015bbz}
\hrefCMSnoop {}{J.~Xu, J.~Liao, and M.~Gyulassy, ``{Bridging soft-hard
  transport properties of quark-gluon plasmas with CUJET3.0}'',} \textit{ JHEP}
  \textbf{ 02} (2016) 169,
  \href{http://dx.doi.org/10.1007/JHEP02(2016)169}{\doi{10.1007/JHEP02(2016)169}},
\href{http://www.arXiv.org/abs/1508.00552}{\texttt{arXiv:1508.00552}}.

\bibitem{Molnar:2003ff}
\hrefCMSnoop {}{D.~Molnar and S.~A. Voloshin, ``{Elliptic flow at large
  transverse momenta from quark coalescence}'',} \textit{ Phys. Rev. Lett.}
  \textbf{ 91} (2003) 092301,
  \href{http://dx.doi.org/10.1103/PhysRevLett.91.092301}{\doi{10.1103/PhysRevLett.91.092301}},
\href{http://www.arXiv.org/abs/nucl-th/0302014}{\texttt{arXiv:nucl-th/0302014}}.

\bibitem{Greco:2003xt}
\hrefCMSnoop {}{V.~Greco, C.~M. Ko, and P.~Levai, ``{Parton coalescence and
  antiproton/pion anomaly at RHIC}'',} \textit{ Phys. Rev. Lett.} \textbf{ 90}
  (2003) 202302,
  \href{http://dx.doi.org/10.1103/PhysRevLett.90.202302}{\doi{10.1103/PhysRevLett.90.202302}},
\href{http://www.arXiv.org/abs/nucl-th/0301093}{\texttt{arXiv:nucl-th/0301093}}.

\bibitem{Fries:2003vb}
\hrefCMSnoop {}{R.~J. Fries, B.~Muller, C.~Nonaka, and S.~A. Bass,
  ``{Hadronization in heavy ion collisions: Recombination and fragmentation of
  partons}'',} \textit{ Phys. Rev. Lett.} \textbf{ 90} (2003) 202303,
  \href{http://dx.doi.org/10.1103/PhysRevLett.90.202303}{\doi{10.1103/PhysRevLett.90.202303}},
\href{http://www.arXiv.org/abs/nucl-th/0301087}{\texttt{arXiv:nucl-th/0301087}}.

\bibitem{Abelev:2007qg}
\hrefCMSnoop {}{{STAR} Collaboration, ``{Mass, quark-number, and
  $\sqrt{s_{NN}}$ dependence of the second and fourth flow harmonics in
  ultra-relativistic nucleus-nucleus collisions}'',} \textit{ Phys. Rev. C}
  \textbf{ 75} (2007) 054906,
  \href{http://dx.doi.org/10.1103/PhysRevC.75.054906}{\doi{10.1103/PhysRevC.75.054906}},
\href{http://www.arXiv.org/abs/nucl-ex/0701010}{\texttt{arXiv:nucl-ex/0701010}}.

\bibitem{Adare:2006ti}
\hrefCMSnoop {}{{PHENIX} Collaboration, ``{Scaling properties of azimuthal
  anisotropy in Au+Au and Cu+Cu collisions at \sqrtsNN\ = 200 GeV}'',} \textit{
  Phys. Rev. Lett.} \textbf{ 98} (2007) 162301,
  \href{http://dx.doi.org/10.1103/PhysRevLett.98.162301}{\doi{10.1103/PhysRevLett.98.162301}},
\href{http://www.arXiv.org/abs/nucl-ex/0608033}{\texttt{arXiv:nucl-ex/0608033}}.

\end{thebibliography}\endgroup

\cleardoublepage \appendix\section{The CMS Collaboration \label{app:collab}}\begin{sloppypar}\hyphenpenalty=5000\widowpenalty=500\clubpenalty=5000\vskip\cmsinstskip
\textbf{Yerevan Physics Institute, Yerevan, Armenia}\\*[0pt]
A.M.~Sirunyan, A.~Tumasyan
\vskip\cmsinstskip
\textbf{Institut f\"{u}r Hochenergiephysik, Wien, Austria}\\*[0pt]
W.~Adam, F.~Ambrogi, E.~Asilar, T.~Bergauer, J.~Brandstetter, E.~Brondolin, M.~Dragicevic, J.~Er\"{o}, A.~Escalante~Del~Valle, M.~Flechl, R.~Fr\"{u}hwirth\cmsAuthorMark{1}, V.M.~Ghete, J.~Hrubec, M.~Jeitler\cmsAuthorMark{1}, N.~Krammer, I.~Kr\"{a}tschmer, D.~Liko, T.~Madlener, I.~Mikulec, N.~Rad, H.~Rohringer, J.~Schieck\cmsAuthorMark{1}, R.~Sch\"{o}fbeck, M.~Spanring, D.~Spitzbart, A.~Taurok, W.~Waltenberger, J.~Wittmann, C.-E.~Wulz\cmsAuthorMark{1}, M.~Zarucki
% ...
\vskip\cmsinstskip
% ...
1:  Also at Vienna University of Technology, Vienna, Austria\\
% ...
\end{sloppypar}
\end{document}